%% file: main.tex
\documentclass[sigconf]{acmart}
\input{packages/packages}
\input{packages/table-packages}

\copyrightyear{2022}
\acmYear{2022}
\setcopyright{acmlicensed}\acmConference[CIKM '22]{Proceedings of the 31st
ACM International Conference on Information and Knowledge
Management}{October 17--21, 2022}{Atlanta, GA, USA}
\acmBooktitle{Proceedings of the 31st ACM International Conference on Information and Knowledge Management (CIKM '22), October 17--21, 2022,
Atlanta, GA, USA}
\acmPrice{15.00}
\acmDOI{10.1145/3511808.3557533}
\acmISBN{978-1-4503-9236-5/22/10}

\begin{document}
\title{MalNet: A Large-Scale Image Database of Malicious Software}

\author{Scott Freitas$^*$}
\email{safreita1@gmail.com}
\affiliation{%
  \institution{Georgia Institute of Technology}
  \city{Atlanta}
  \state{GA}
  \country{USA}
}

\author{Rahul Duggal}
\email{rahulduggal@gatech.edu}
\affiliation{%
  \institution{Georgia Institute of Technology}
  \city{Atlanta}
  \state{Georgia}
  \country{USA}
}

\author{Duen Horng Chau}
\email{polo@gatech.edu}
\affiliation{%
  \institution{Georgia Institute of Technology}
  \city{Atlanta}
  \state{Georgia}
  \country{USA}
}

\thanks{* Now at Microsoft, work completed when at Georgia Tech.}

\renewcommand{\shortauthors}{Scott Freitas, Rahul Duggal, \& Duen Horng Chau}

\begin{abstract}
Computer vision is playing an increasingly important role in automated malware detection with the rise of the image-based binary representation.
These binary images are fast to generate, require no feature engineering, and are resilient to popular obfuscation methods.
Significant research has been conducted in this area, however, it has been restricted to small-scale or private datasets that only a few industry labs and research teams have access to.
This lack of availability hinders examination of existing work, development of new research, and dissemination of ideas.
We release \dataset, the largest public cybersecurity image database, offering 24$\times$ more images and 70$\times$ more classes than existing databases (available at {\small\url{https://mal-net.org}}).
\dataset contains over 1.2 million malware images---across 47 types and 696 families---democratizing image-based malware capabilities by enabling researchers and practitioners to evaluate techniques that were previously reported in propriety settings.
We report the first million-scale malware detection results on binary images.
\dataset unlocks new and unique opportunities to advance the frontiers of machine learning, enabling new research directions into vision-based cyber defenses, multi-class imbalanced classification, and interpretable security. 
\end{abstract}


\maketitle

\input{sections/01-introduction}

\input{sections/02-construction}

\input{sections/03-applications}

\input{sections/04-conclusion}


\begin{acks}
We thank Kevin Allix and AndroZoo colleagues for generously allowing us to use their data in this research; this work was in part supported by NSF grant IIS-1563816, CNS-1704701, GRFP (DGE-1650044), a Raytheon research fellowship, and an IBM fellowship.
\end{acks}

\bibliographystyle{ACM-Reference-Format}
\bibliography{main.bib}


\end{document}

%% file: packages/packages.tex
\usepackage{times}
\usepackage{graphicx}
\usepackage{amsmath}
\usepackage{xspace}
\usepackage{enumitem}

\newif\ifcomment

\commenttrue

\newcommand{\dataset}{\textsc{MalNet-Image}\xspace}

%% file: packages/table-packages.tex
\usepackage{xcolor}
\usepackage{multirow}
\usepackage{colortbl}
\usepackage{arydshln}

\definecolor{header}{HTML}{db2525} 
\definecolor{identifier}{HTML}{5079cc} 
\definecolor{data}{HTML}{175fd4}
\definecolor{lightgray}{HTML}{363636}

%% file: sections/01-introduction.tex
\section{Introduction}

\input{tables/comparison}

Attack campaigns from criminal organizations and nation state actors are one of the most powerful forms of disruption, costing the U.S. economy as much as \$109 billion a year~\cite{council2018}. 
These cyber attacks are highly sophisticated, targeting governments and large-scale enterprises to interrupt critical services and steal intellectual property~\cite{freitas2020d2m}.
Defending against these attacks requires the development of strong antivirus tools to identify new variants of malicious software before they can infect a network.
Unfortunately, as a majority of newly identified malware is \textit{polymorphic} in nature, where a few subtle source code changes result in significantly different compiled code (e.g., instruction reordering, branch inversion, register allocation)~\cite{dullien2005graph,you2010malware}, the predominant signature-based form of malware detection is rendered inert~\cite{sathyanarayan2008signature}.

To combat these issues, the cybersecurity industry~\cite{chen2020stamina} has turned to image-based malware representations as they are quick to generate, require no feature engineering, and are resilient to common obfuscation techniques (e.g., section encryption~\cite{nataraj2011malware}, file packing~\cite{nataraj2011comparative}).
For all of these reasons, image-based malware detection and classification research has surged in popularity.
Unfortunately, a majority of this research uses small-scale or private data repositories, making it increasingly difficult to characterize and differentiate existing work, develop new research methodologies, and disseminate new ideas~\cite{chen2020stamina,conti2010visual,fang2020android,fu2018malware,gennissen2017gamut,han2015malware,lu2019new,luo2017binary,nataraj2011malware,nataraj2011comparative,raff2018malware}.
To address these issues, we constructed \dataset, the first large-scale ontology of malicious software images. 

\subsection{Contributions}

\noindent\textbf{1. Largest Cybersecurity Image Database.}
\dataset contains over 1.2 million software images across a hierarchy of $47$ types and $696$ families, enabling researchers and practitioners to conduct experiments on an industry scale dataset, and evaluate techniques that were previously reported in propriety settings.
Compared to the next large public database~\cite{noever2021virus}, \dataset offers $24\times$ more images and nearly $70\times$ more classes (see Table~\ref{table:dataset_comparison}).
We report the first public large-scale malware detection and classification results on binary images, where we are able to detect malicious files with an AUC of $0.94$ and classify them across $47$ types and $696$ families with a macro-F1 score of $0.49$ and $0.45$, respectively.
    
\medskip\noindent\textbf{2. Permissive Licensing \& Open Source Code.}
We release \dataset with a CC-BY license, allowing researchers and practitioners to share and adapt the database to their needs.
We open-source the code to create the images and run the experiments on \href{https://github.com/safreita1/malnet-image}{Github}.

\medskip\noindent\textbf{3. Visual Exploration Without Downloading.}
We develop \textsc{\dataset{} Explorer}, an image exploration and visualization tool that enables researchers and practitioners to easily study the data without installation or download.
\textsc{\dataset{} Explorer} is available online at: \url{https://mal-net.org}.

\medskip\noindent\textbf{4. Community Impact.}
\dataset offers new and unique opportunities to advance the frontiers of cybersecurity research.
In particular, \dataset offers researchers a chance to study imbalanced classification on a large-scale cybersecurity database with a natural imbalance ratio of $16,901\times$ (see Figure~\ref{fig:imbalance}); and explore explainability research in a high impact domain, where it is critical that security analysts can interpret and trust the model.

%% file: tables/comparison.tex
\begin{table}
\vspace{6mm}
\setlength{\tabcolsep}{13pt}
\renewcommand{\arraystretch}{1.3}
\centering
\begin{tabular}{llrr}
\toprule
 
& Dataset & Images & Classes \\

\cmidrule(l){2-2} \cmidrule(l){3-3} \cmidrule(l){4-4}
\multirow{3}{*}{\textcolor{data}{{\textbf{\rotatebox[origin=c]{90}{Public}}}}}
& \small{\textcolor{data}{\textbf{\dataset}}} & \textcolor{data}{\textbf{1,262,024}} & \textcolor{data}{\textbf{696}} \\
& \textcolor{lightgray}{Virus-MNIST~\cite{noever2021virus}} & \textcolor{lightgray}{51,880} & \textcolor{lightgray}{10} \\
& \textcolor{lightgray}{Malimg~\cite{nataraj2011comparative}} & \textcolor{lightgray}{9,458} & \textcolor{lightgray}{25} \\\\[-0.85em]

\cdashline{1-4}[1pt/1.5pt]\noalign{\vskip 0.65em}

\multirow{7}{*}{{\textcolor{header}{\textbf{\rotatebox[origin=c]{90}{Private}}}}}
& \textcolor{lightgray}{Stamina~\cite{chen2020stamina}} & \textcolor{lightgray}{782,224} & \textcolor{lightgray}{2} \\
& \textcolor{lightgray}{McAfee~\cite{gennissen2017gamut}} & \textcolor{lightgray}{367,183} & \textcolor{lightgray}{2} \\
& \textcolor{lightgray}{Kancherla~\cite{kancherla2013image}} & \textcolor{lightgray}{27,000} & \textcolor{lightgray}{2} \\
& \textcolor{lightgray}{Choi~\cite{choi2017malware}} & \textcolor{lightgray}{12,000} & \textcolor{lightgray}{2} \\
& \textcolor{lightgray}{Fu~\cite{fu2018malware}} & \textcolor{lightgray}{7,087} & \textcolor{lightgray}{15} \\
& \textcolor{lightgray}{Han~\cite{han2015malware}} & \textcolor{lightgray}{1,000} & \textcolor{lightgray}{50} \\
& \textcolor{lightgray}{IoT DDoS~\cite{su2018lightweight}} & \textcolor{lightgray}{365} & \textcolor{lightgray}{3} \\

\bottomrule
\end{tabular}
\caption{\dataset: a state-of-the-art cybersecurity image
database containing over 1.2 million binary images across
a hierarchy of 47 types and 696 families. 
}
\vspace{-8mm}
\label{table:dataset_comparison}
\end{table}

%% file: sections/02-construction.tex
\section{Advancing the State-of-the-Art}
Aside from \dataset, there are only two publicly available binary-image based cybersecurity datasets--- {Malimg}~\cite{nataraj2011comparative} and Virus-MNIST~\cite{noever2021virus}---containing 9,458 images across 25 classes, and 51,880 images across 10 classes, respectively. 
In surveying the malware detection and classification literature \cite{nataraj2011comparative,chen2020stamina,gennissen2017gamut,kancherla2013image,choi2017malware,fu2018malware,han2015malware,su2018lightweight,mclaughlin2017deep,mercaldo2020deep,burks2019data,azab2020msic,yue2017imbalanced,catak2020data,ren2020end,chen2018deep,luo2017binary,jain2015enriching,kumar2016machine,fang2020android}, we observed that almost all experiments were conducted on small-scale or private data.
As the field advances, large-scale public databases are necessary to develop the next generation of algorithms.
In Table~\ref{table:dataset_comparison}, we compare \dataset with other public and private cybersecurity image datasets.
We find that that \dataset offers 24$\times$ more images and 70$\times$ the classes, compared to the largest alternative public binary image database (Virus-MNIST~\cite{noever2021virus}); and $479,800$ more images and $694$ more classes than the largest private database (Stamina~\cite{chen2020stamina}). 
We do not compare against repositories of malicious binaries such as AndroZoo~\cite{li2017androzoo++}, AMD~\cite{wei2017deep}, Microsoft-BIG~\cite{ronen2018microsoft},  Malicia~\cite{nappa2013driving}, VirusShare, and VirusTotal in this discussion, as none of them have images available to use.

\medskip
\noindent\textbf{Security Implications.}
With the release of \dataset, researchers will now have access to a critical resource to develop advanced, image-based malware detection and classification algorithms.
Like most open data resources, there is a potential for misuse by malicious actors who aim to craft new variants to evade detection.
We believe \dataset's contribution to the research community significantly outweighs such risk.

\begin{figure}[t]
\centering
\includegraphics[width=\linewidth]{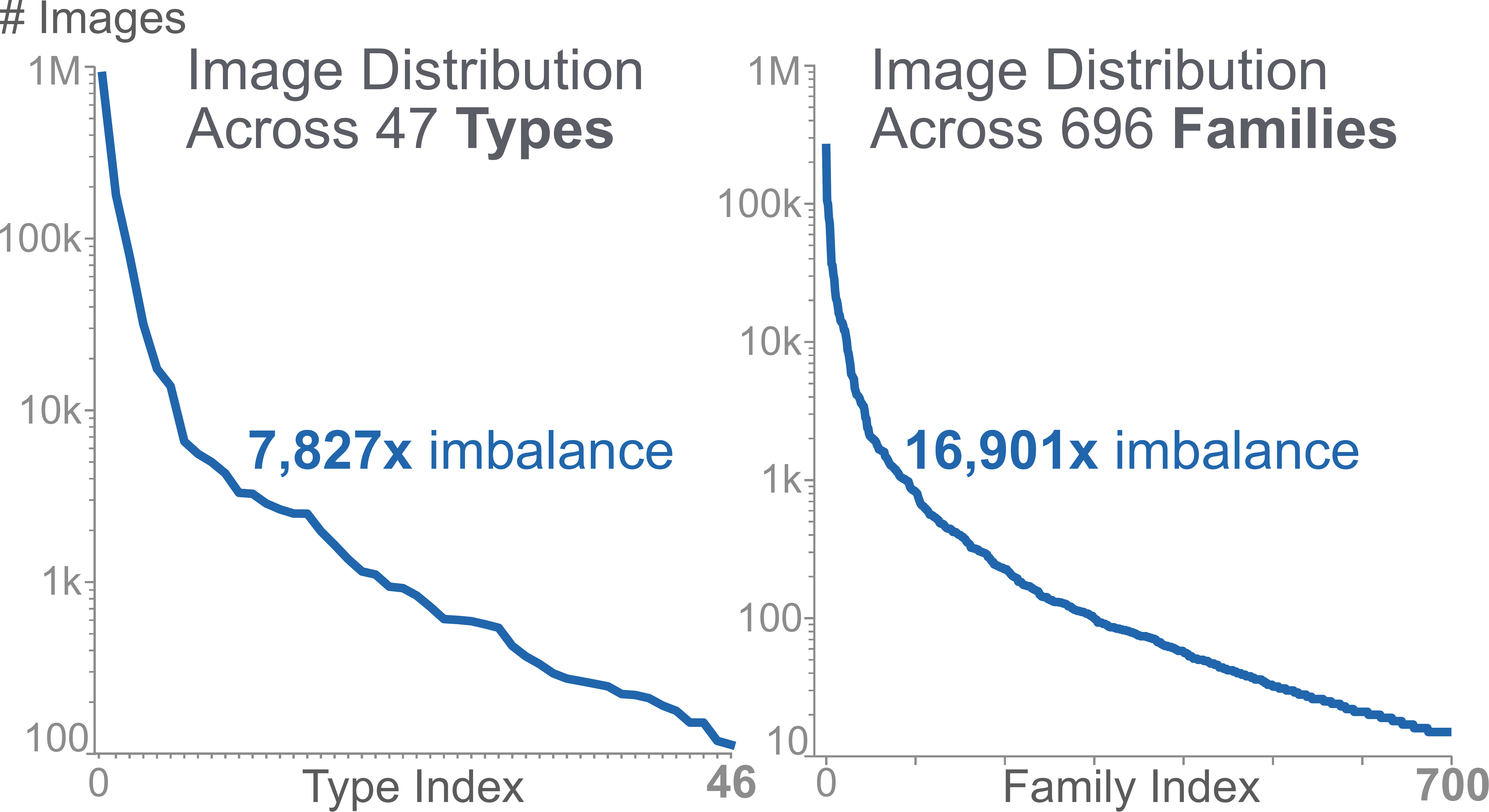}
\vspace{-6mm}
\caption{Class imbalance distribution for \textit{type} and \textit{family}.
}
\label{fig:imbalance}
\end{figure}

\subsection{Constructing MalNet-Image}\label{sec:construction}
\dataset is an ambitious project to collect and process over 1.2 million binary images, and is a major extension to the graph representation learning database \textsc{MalNet}~\cite{freitas2020large}, offering significant new malware detection capabilities.
Below, we describe the provenance and construction of \dataset.

\medskip
\noindent\textbf{Collecting Candidate Images.}
We construct \dataset using the Android ecosystem due to its large market share~\cite{popper2017}, easy accessibility~\cite{li2017androzoo++} and diversity of malicious software~\cite{nokia2019}.
With the generous permission of AndroZoo~\cite{allix2016androzoo,li2017androzoo++}, we collected 1,262,024 Android APK files, specifically selecting APKs containing both a \textit{family} and \textit{type} label obtained from Euphony~\cite{hurier2017euphony}, a state-of-the-art malware labeling system that aggregates and learns from the labelling results of up to 70 antivirus vendors from VirusTotal~\cite{total2012virustotal}.

\begin{figure}[b]
\centering
\includegraphics[width=\linewidth]{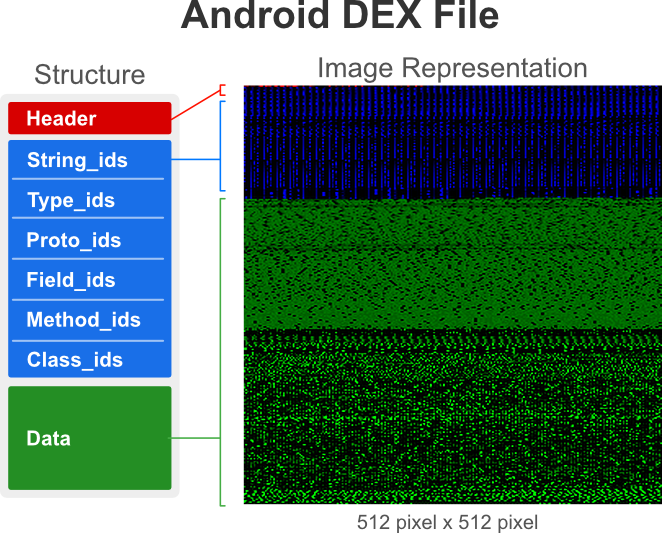}
\caption{\textbf{Left:} Android DEX file structure, composed of three major components---(1) header, (2) ids, and (3) data.
\textbf{Right:} binary image representation of the DEX file.
}
\label{fig:dex-structure}
\end{figure}

\medskip
\noindent\textbf{Processing the Images.}
The first step in constructing the image representation was to extract the DEX file (bytecode) from each Android APK.
The extracted DEX file was then converted into a 1D array of $8$ bit unsigned integers.
Each entry in the array is in the range $[0, 255]$ where $0$ corresponds to a black pixel and $255$ a white pixel.
We then convert each 1D byte array into a 2D array using standard linear plotting where the width of the image is fixed and the height is allowed to vary based on the file size.
We use the same image width proposed in the seminal work \cite{nataraj2011malware} (and follow up work~\cite{kalash2018malware,cui2018detection,rezende2018malicious,yakura2018malware}), and scale each image to $256\times 256$ using a standard Lanczos filter from the Pillow library.
Finally, we color each byte according to its use, adding a layer of semantic information on top of the raw bytecode.
While a variety of techniques can be used to encode semantic information into the image, there is currently no accepted standard.
We follow \cite{gennissen2017gamut} and assign each byte to a particular RGB channel depending on its position in the DEX file structure---(i) \textit{header}, (ii) \textit{identifiers} and \textit{class definitions}, and (iii) \textit{data} (see Figure~\ref{fig:dex-structure}).
Distributed across Google Cloud's General-purpose (N2) machine with $16$ cores running $24$ hours a day, this process took approximately a week.
We release the source code used to process the APKs on \href{https://github.com/safreita1/malnet-image}{Github}.

\medskip
\noindent\textbf{MalNet-Image Tiny.}
We construct \textsc{\dataset Tiny}, containing $61,201$ training, $8,743$ validation and $17,486$ test images, for \textit{type level classification} experiments by removing the 4 largest types in \dataset.
The goal of \textsc{\dataset-Tiny} is to enable users to rapidly prototype new ideas, since it requires only a fraction of the time needed to train a new model.
\textsc{\dataset Tiny} is released alongside the full dataset at {\small\url{https://mal-net.org}}.

%% file: sections/03-applications.tex
\section{MalNet-Image Applications}\label{sec:application}
\dataset offers new and unique opportunities to advance the frontiers of cybersecurity research.
As examples, we show three exciting new applications made possible by the \dataset database---(1) as a state-of-the-art cybersecurity image benchmark in Section~\ref{subsec:app_baselines};
(2) as the first large-scale public analysis of malicious software detection using binary images in Section~\ref{subsec:malware-detection};
and (3) how to categorize high-risk malware threats (e.g., is this Ransomware or Spyware?) in Section~\ref{subsec:malware-classification}.
Then, in Section~\ref{subsec:research-challenges} we highlight new research directions enabled by \dataset.

\medskip\noindent
\textbf{Application Setup.}
We divide \dataset into three stratified sets of data, with a training-validation-test split of $70$-$10$-$20$ respectively; 
repeated for both type and family labels (suggested splits available at {\small\url{https://mal-net.org}}).
In addition, we conduct malware detection experiments by grouping all 46 malicious software images into one type while the benign type maintains its original label.
We evaluate 3 common architectures---ResNet~\cite{he2016deep}, DenseNet~\cite{huang2017densely} and MobileNet~\cite{howard2017mobilenets}, based on its macro-F1 score, as is typical for highly imbalanced datasets~\cite{duggal2020elf,duggal2020rest,duggal2021har,freitas2020large}.
Each model is trained for 100 epochs using cross entropy loss  (unless specified otherwise) and an Adam optimizer on an Nvidia DGX-1 containing 8 V100 GPUs and 512GB of RAM using Keras with a Tensorflow backend.

\subsection{Application 1: Benchmarking Techniques}\label{subsec:app_baselines}
Leveraging the unprecedented scale and diversity of \dataset, we evaluate numerous malware detection and classification techniques that have previously been studied using only private or small-scale databases.
Specifically, we evaluate recent techniques including: (a) semantic information encoding via colored channels, (b) model architecture, (c) imbalanced classification techniques, and (d) the performance of \textsc{\dataset Tiny}, a small-scale version of \dataset.
We detail the setup, results, and analysis of each experiment below.

\medskip\noindent
\textbf{Semantic Information Encoding.}
We evaluate the effect of information encoding in the classification process by training two ResNet18 models---one on the RGB images, where each byte is assigned to a particular color channel depending on its position in the DEX file structure as proposed in \cite{gennissen2017gamut}, and another on grayscale converted images.
We find no improvement in the macro-F1 score using semantically encoded RGB images compared to grayscale ones. 
As there are alternative encoding techniques~\cite{gennissen2017gamut}, we believe comparing the effects of different encodings could be an interesting  future research direction.
Going forward, all models are trained using grayscale images.

\input{tables/results-models}

\medskip\noindent
\textbf{Evaluating Model Architectures.} 
We evaluate malware detection and classification performance on 3 popular deep learning architectures (ResNet, DenseNet and MobileNetV2) across a variety of model sizes, using grayscale encoded images, and cross entropy loss.
In Table~\ref{tab:large_scale_results}, we report the macro-F1, macro-precision, and macro-recall of each model.
We find that all models obtain similar macro-F1 scores, indicating that a small model has enough capacity to learn the features present in the binary images.
Going forward, all experiments use a ResNet18 model due to its strong performance and fast training time.

\medskip\noindent
\textbf{Accounting for Class Imbalance.}
We evaluate 3 imbalanced classification techniques---(1) class reweighting with cross entropy loss, (2) focal loss, and (3) class reweighting with focal loss; and compare this to a model trained using cross entropy loss without class weighting. 
For class reweighting, each example of a class $c$ is weighted according to it's effective number $\frac{1-\beta}{1-\beta^{n_c}}$, where $n_c$ is the number of images in class $c$ and $\beta=0.999$ is selected through a line search across standard values~\cite{cui2019class} of $\{0.9, 0.99, 0.999, 0.9999\}$.
For focal loss~\cite{lin2017focal}, a regularization technique that tackles imbalance by establishing margins based on the class size, we set the hyperparameter $\gamma=2$ as suggested in \cite{lin2017focal}.

Analyzing the results, we find that cross entropy loss with class reweighting improves the \textit{type} macro-F1 score by $0.021$, but lowers the binary and family classification scores by $0.002$ and $0.006$, respectively.
In particular, we notice that \dataset's smallest types benefit the most from class reweighting, where the `Click' type (113 examples), sees its F1 score rise from $0$ to $0.91$.
On the other hand, focal loss shows no improvement over the baseline model, likely due to its design for use in dense object detectors like R-CNN.
Going forward, all experiments use cross entropy loss with class reweighting due to the strong improvement in smaller malware types.

\medskip\noindent
\textbf{\textsc{MalNet-Image Tiny} Performance.} 
We train a ResNet18 model on grayscale images using cross entropy loss and class reweighting, and achieve a macro-F1 score of $0.65$. 
Compared to full dataset, the macro-F1 score is significantly higher $0.65$ vs $0.49$; which is unsurprising since the largest 4 types contained a significant proportion of the image diversity (based on the number of families), resulting in an easier classification task.

\medskip\noindent
\textbf{Limitations.} Methods that work well on other datasets may not work well on \dataset due to structural differences in the images; vice-versa, methods that work on \dataset may not transfer well to other datasets.
We hope this work inspires new research in the binary image domain, enabling the development of methods that generalize across key domains such as cybersecurity.

\subsection{Application 2: Malware Detection}\label{subsec:malware-detection}

Researchers and practitioners can now conduct malware detection experiments on an industry scale dataset, evaluating things that were previously reported in propriety settings.
Using the model selected in Section~\ref{subsec:app_baselines}---a ResNet18 model trained on grayscale images using cross entropy loss and class reweighting---we perform an in-depth analysis of this highly imbalanced detection problem containing $1,182,905$ malicious and $79,119$ benign images.
We find that the model is able to obtain a strong macro-F1 score of $0.86$, macro-precision of $0.89$ and a macro-recall of $0.84$.
We further study the model's detection capabilities by analyzing its ROC curve, where the model achieves an AUC score of $0.94$, and is able to identify $84\%$ of all malicious files with a false positive rate of $10\%$ (a common threshold used in security~\cite{chen2020stamina}).
This first of its kind analysis allows researchers insight into malware detection that is usually restricted to handful of industry labs.

\begin{figure}
\centering
\includegraphics[width=\linewidth]{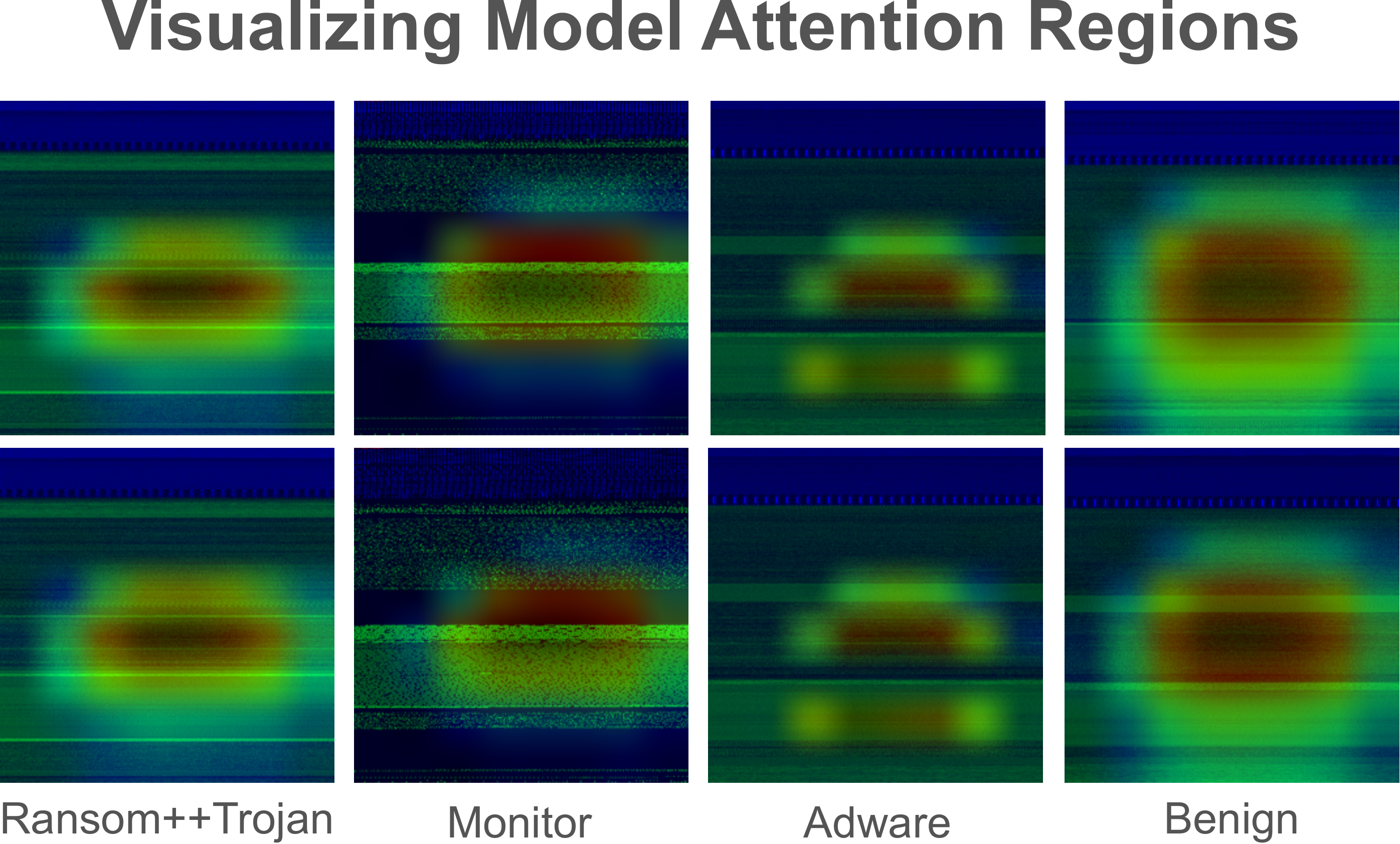}
\caption{Model attention patterns across $4$ malware types (each with $2$ images).
\textbf{Ransom++Trojan:} focus on thin region of data section.
\textbf{Benign:} wide range of attention across data section.
\textbf{Adware:} attention on circular bytecode ``hotspots''.
\textbf{Monitor:} focus on ``empty''  black region of data section.
}
\label{fig:model-attention}
\end{figure}

\dataset also opens new opportunities in the nascent and promising research direction of analyzing attention maps to interpret malware detection results.
Yakura et al.~\cite{yakura2019neural,yakura2018malware} showed that specific byte sequences found in the attention map closely correlate with malicious code payloads.
We evaluate the potential of attention maps on \dataset using the popular Grad-Cam~\cite{selvaraju2017grad} technique to highlight regions of interest across 3 types of malware and benignware in Figure~\ref{fig:model-attention}.
In the malware images (left three), we see the attention map is focused on thin regions of bytecode in the data section (where malicious payloads are often stored), while in the benign images (right side) the attention map is dispersed across the larger data region.
This type of visual analysis can significantly reduce the amount of time and effort required to manually investigate a file by guiding security analysts to suspicious regions of the bytecode.

\subsection{Application 3: Malware Classification}\label{subsec:malware-classification}
\dataset opens up new research into binary images as a tool for multi-class malware classification (e.g., is this file Ransomware or Spyware?).
Using the model selected in Section~\ref{subsec:app_baselines}, we perform an in-depth analysis of its multi-class classification capability across $47$ types and $696$ families of malware.
We find the model is able to classify the malware \textit{type} and malware \textit{family} with a macro-F1 score of $0.49$ and $0.45$, respectively.

\begin{figure*}
\includegraphics[width=\textwidth]{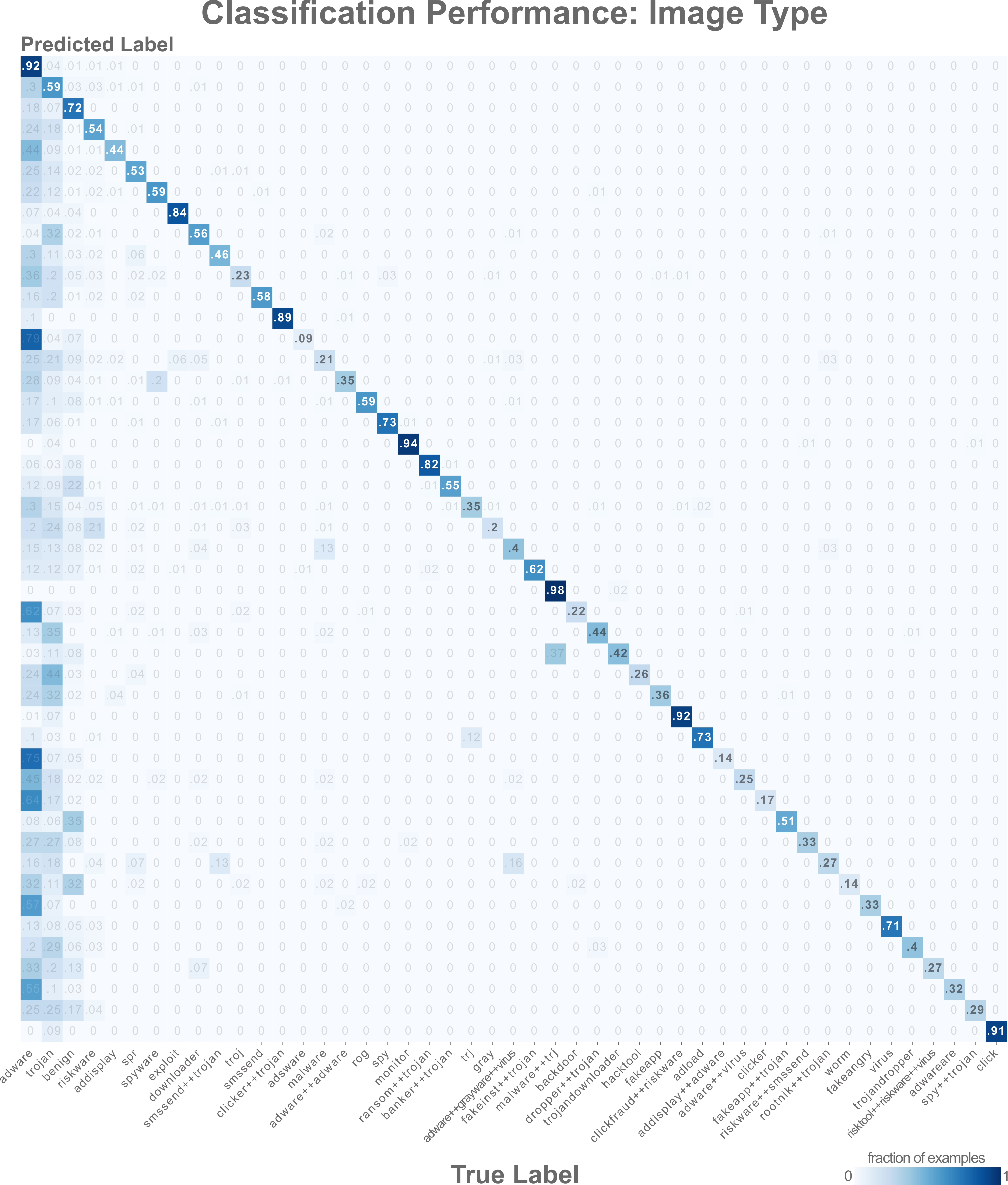}
\caption{
Malware classification results using confusion matrix heatmap (classes in descending order of number of samples).
We analyze type level classification performance, where a dark diagonal indicates strong performance, and a dark off-diagonal indicates poor performance. 
Each square in the diagonal indicates the percent of examples correctly classified for a particular malware type, and each off-diagonal entry indicates the percent of incorrectly classified examples for a particular type.
}
\label{fig:cm}
\end{figure*}

In Figure~\ref{fig:cm}, we conduct an in-depth analysis into \textit{type} level classification performance through a confusion matrix heatmap.
A dark diagonal indicates strong classifier performance, where a dark off-diagonal entry indicates poor performance. 
Each square in the diagonal indicates the percent of examples correctly classified for a particular malware type, and each off-diagonal row entry indicates the percent of incorrectly classified examples for a particular malware type.
We find that four types of malware comprise the majority of misclassifications: Adware, Benign, Riskware, and Trojan.
Unsurprisingly, these are the 4 largest types of malware (based on the number of images in each class), indicating the strong effect that data imbalance has in the malware classification process.
In addition, the heatmap can be used to identify potential naming disagreements between vendor labels (e.g., ``adware'' and ``adsware''), serving as the basis for merging certain types of malware.
To the best of our knowledge, this is the first public large-scale analysis of malware classification, providing a new state-of-the-art benchmark to compare against.

\subsection{Enabling New Research Directions}\label{subsec:research-challenges}
The scale and diversity of \dataset opens up new exciting research opportunities to the ML and security communities. Below, we present 3 promising directions (R1-R3).

\begin{enumerate}[topsep=2mm, itemsep=1mm, parsep=1mm, leftmargin=*, label=\textbf{R\arabic*.}]

\item \textbf{Advancing Vision Based Cybersecurity Research.} 
Research into developing image-based malware detection and classification algorithms has recently surged across industry (e.g., Intel-Microsoft collaboration on Stamina~\cite{chen2020stamina}, security companies~\cite{noever2021virus,gennissen2017gamut}) and academia~\cite{kancherla2013image,choi2017malware,fu2018malware,han2015malware,su2018lightweight,mclaughlin2017deep,mercaldo2020deep,burks2019data,azab2020msic,yue2017imbalanced,catak2020data,ren2020end,chen2018deep,luo2017binary,jain2015enriching,kumar2016machine,fang2020android}.
However, existing public datasets contain only a handful of classes and thousands of images, and as the field advances, larger and more challenging datasets are needed for the next generation of models.
With the release of \dataset, researchers have access to a critical resource to develop and benchmark advanced image-based malware detection and classification algorithms, previously restricted to a few industry labs and research teams.

\item \textbf{Extending Imbalanced Classification to a New Domain.}
Only preliminary work has studied binary-image malware classification under data imbalance~\cite{yue2017imbalanced} due to the limited number of classes and images available in public datasets.
As a result, it is unknown whether many techniques may generalize to the binary-image domain, and how they will perform in highly imbalanced classification scenarios.
We take a first step in this direction in Section~\ref{subsec:app_baselines}, where we show that classes containing only a few examples typically underperform relative to their more populous counterparts---highlighting the significant challenge of imbalanced classification in the cybersecurity domain. 
By releasing \dataset, one of the largest naturally imbalanced databases to date, we hope to foster new interest in this important research area, enabling the machine learning community to impact and generalize across domains.

\item \textbf{Interpretable Cybersecurity Research.} 
Preliminary research has demonstrated the importance of attention mechanisms in binary-image malware classification, where extracted regions can provide strong indicators to human analysts, helping guide them to suspicious parts of the bytecode for additional analysis~\cite{yakura2019neural,yakura2018malware}.
This includes recent research in salience based methods that automatically discover concepts, helping to identify correlated regions of bytecode~\cite{zhao2015saliency}.
Prior to \dataset, researchers were limited to a small number of malicious families and types, hindering their ability to conduct large-scale explainability studies.
With \dataset's nearly 700 classes, researchers can explore a wide variety of malicious software, enabling new breakthroughs and discoveries.
For example, researchers might discover that new types of visualization and sense-making techniques are needed to accurately summarize large volumes of binary-image data to enhance security analysts decision making capabilities. 
\end{enumerate}

%% file: tables/results-models.tex
\begin{table}[t]
\centering
\small
\renewcommand{\arraystretch}{1.5}
\setlength{\tabcolsep}{3.4pt}
\begin{tabular}{lrrrrrrrrrrr}

\toprule
& & & \multicolumn{3}{c}{\textbf{Binary}} & \multicolumn{3}{c}{ \textbf{Type}} & \multicolumn{3}{c}{\textbf{Family}} \\ 

\cmidrule(l{5pt}r{5pt}){4-6}\cmidrule(l{5pt}r{5pt}){7-9}\cmidrule(l{5pt}r{5pt}){10-12}

\textbf{Model} & \textbf{Param} & \textbf{GFlop} & F1 & P & R &  F1 & P & R & F1 & P & R \\
\hline

RN18 & 12M & 1.8 & .86 & .89 & .84 & .47 & .56 & .42 & .45 & .54 & .42 \\
RN50 & 26M & 3.9 & .85 & .91 & .81 & .48 & .57 & .44 & .47 & .54 & .44 \\
RN101 & 45M & 7.6 & .86 & .88 & .84 & .48 & .59 & .44 & .47 & .54 & .44 \\

DN121 & 7.9M & 2.9 & .86 & .90 & .83 & .47 & .56 & .43 & .46 & .53 & .44 \\
DN169 & 14M & 3.4 & .86 & .89 & .84 & .48 & .57 & .43 & .46 & .55 & .43 \\

MN2\scriptsize{(x.5)} & 1.9M & 0.1 & .86 & .89 & .83 & .46 & .55 & .42 & .45 & .53 & .42 \\
MN2\scriptsize{(x1)} & 3.5M & 0.3 & .85 & .89 &  .83 & .45 & .53 & .42 & .44 & .53 & .41 \\

\bottomrule
\end{tabular}
\caption{
We evaluate the performance of 3 architectures---ResNet, DenseNet and MobileNetV2---on its macro-F1, macro-precision, and macro-recall.
We conduct the remaining experiments using the ResNet18 model as it provides a strong balance between performance and efficiency.
}
\label{tab:large_scale_results}
\vspace{-5mm}
\end{table}

%% file: sections/04-conclusion.tex
\section{Conclusion}\label{sec:conclusion}
Computer vision research into binary-image malware detection and classification is a crucial tool in protecting enterprise networks and governments from cyber attacks seeking to interrupt critical services and steal intellectual property.
Leveraging \dataset's scale and diversity---containing $1,262,024$ binary images across a hierarchy of $47$ types and $696$ families---researchers and practitioners can now conduct experiments that were previously restricted to a few industry labs and research teams. 
We hope \dataset becomes a central resource for a broad range of research into vision-based cyber defenses, multi-class imbalanced classification, and interpretable security.